\documentclass[10 pt]{amsart}
\usepackage[pdftex]{graphicx}
\usepackage{color}
\newcommand{\edit}[1]{\textcolor{black}{#1}}
\newcommand{\x}{\mathbf{x}}

\begin{document}

\title{Real Time Adaptive Event Detection in Astronomical Data Streams: Lessons from the VLBA}
\author{David R. Thompson$^{1,2}$}
\author{Sarah Burke-Spolaor$^2$}
\author{Adam T. Deller$^{3}$}
\author{Walid A. Majid$^2$}
\author{Divya Palaniswamy$^4$}
\author{Steven J. Tingay$^4$}
\author{Kiri L. Wagstaff$^2$}
\author{Randall B. Wayth$^4$}

\thanks{$^1${\tt david.r.thompson@jpl.nasa.gov}\\
$^2$ Jet Propulsion Laboratory, California Institute of Technology, 4800 Oak Grove Dr., Pasadena, CA 91109, USA\\
$^3$ ASTRON, Oude Hoogeveensedijk 4, 7991 PD Dwingeloo, The Netherlands\\
$^4$ International Centre for Radio Astronomy Research, Curtin University. GPO Box U1987, Perth WA, 6845. Australia}
\newpage
\begin{abstract}  
A new generation of observational science instruments is dramatically increasing collected data volumes in a range of fields.  These instruments include the Square Kilometre Array (SKA), Large Synoptic Survey Telescope (LSST), terrestrial sensor networks, and NASA satellites participating in ``decadal survey'' missions.  Their unprecedented coverage and sensitivity will likely reveal wholly new categories of unexpected and transient events.  {\it Commensal} methods passively analyze these data streams, recognizing anomalous events of scientific interest and reacting in real time.  We report on a case example: V-FASTR, an ongoing commensal experiment at the Very Long Baseline Array (VLBA) that uses online adaptive pattern recognition to search for anomalous {\it fast radio transients}.  V-FASTR triages a millisecond-resolution stream of data and \edit{promotes candidate anomalies for} further offline analysis.   \edit{It tunes detection parameters in real time}, injecting synthetic events to continually retrain itself for optimum performance.  This self-tuning approach retains sensitivity to weak signals while adapting to changing instrument configurations and noise conditions.  The system has operated since July 2011, making it the longest-running real time commensal radio transient experiment to date.  

\vspace{0.3cm}
\end{abstract}

\maketitle
\thispagestyle{empty}

\section{Introduction}

The next generation of scientific instruments will dramatically increase collected data volumes in multiple disciplines.   Astronomers will soon face petascale data streams from instruments like the Large Synoptic Survey Telescope and Square Kilometer Array \cite{Dewdney2009}.   The environmental remote sensing community will see orders of magnitude increases in data volume from ``decadal survey'' satellite missions such as OCO-2, HyspIRI and GeoCAPE.  These instruments will almost certainly observe secondary transient events and anomalies of scientific interest alongside their primary missions.  Examples of transient activities include orbiters that serendipitously observe wildfires or erupting volcanoes and ground-based telescopes that observe supernovae and other momentary astronomical events.  \edit{These generally require immediate action}, either to trigger additional observations or to archive contextual data \edit{needed for} validation.  The time window for followup action is typically quite short, demanding efficient real time detection algorithms.  

This work describes \edit{a case study investigating} {\it fast radio transients}.  These millisecond-scale pulses of radio frequency energy may be produced by astronomical events such as Gamma Ray Bursts, intermittent radio pulsars known as RRATs, or possibly more extreme events such as merging black holes \cite{Cordes2004,Lazio2009}.  At the time of this writing, few candidate transients have been discovered in past archives \cite{Keane2012}, and there have been no \edit{wholly conclusive detections} of the most exotic events.  More examples are needed to validate these detections and characterize the population.  To this end, passive {\it commensal analyses}, which operate in piggyback on other investigations, are an appealing way to accumulate significant telescope time.

Radio transient detection typifies common challenges of large-scale commensal \edit{searches}.  One is consistency.  Commensal analysis must cope with a range of different interference conditions and configurations. It must also keep pace with any instrument upgrades and modifications.  \edit{Another challenge is data volume.}  The system operates continuously with high duty cycle and processing loads.  This requires an automated system that can triage the data stream, promoting only the most important segments to storage and analyst attention.  

Here we describe the Very Long Baseline Array Fast Transients Experiment (V-FASTR), a system for commensal detection of fast radio transient events \cite{Wayth2011,Thompson2011}.  V-FASTR is a pathfinder project for a larger transient detection system known as CRAFT. \edit{It searches data from the Very Long Baseline Array (VLBA) for transient events and anomalies.  Previous work analyzed event detection algorithms using simulation and controlled tests prior to deployment \cite{Thompson2011}.  Here we report on the actual performance during the first year of operations.  We also describe a novel new approach for {\it self-tuning} of detection parameters by injecting synthetic events into the data stream.  This adapts to a range of instrument conditions, detecting rare, subtle astronomical events while retaining resilience to noise.  As of this writing, V-FASTR is the single longest running commensal transient search to date and an instructive example of real time online event detection.} 

\section{The V-FASTR System}

The Very Long Baseline Array (VLBA) consists of ten identical antennas at locations across the western hemisphere \cite{Romney2010}.   Typical VLBA operations point all of the telescopes at a common location to perform interferometric imaging with very high spatial resolution.  Each antenna records raw voltage samples that are sent in physical disks to the array operations center in Socorro, NM.  Here they are time-aligned, {\it correlated}, and processed to form an interferometric radio sky image.  After this step the disks are erased for reuse.

\begin{figure}[htb]
\includegraphics[width=\textwidth,angle=0]{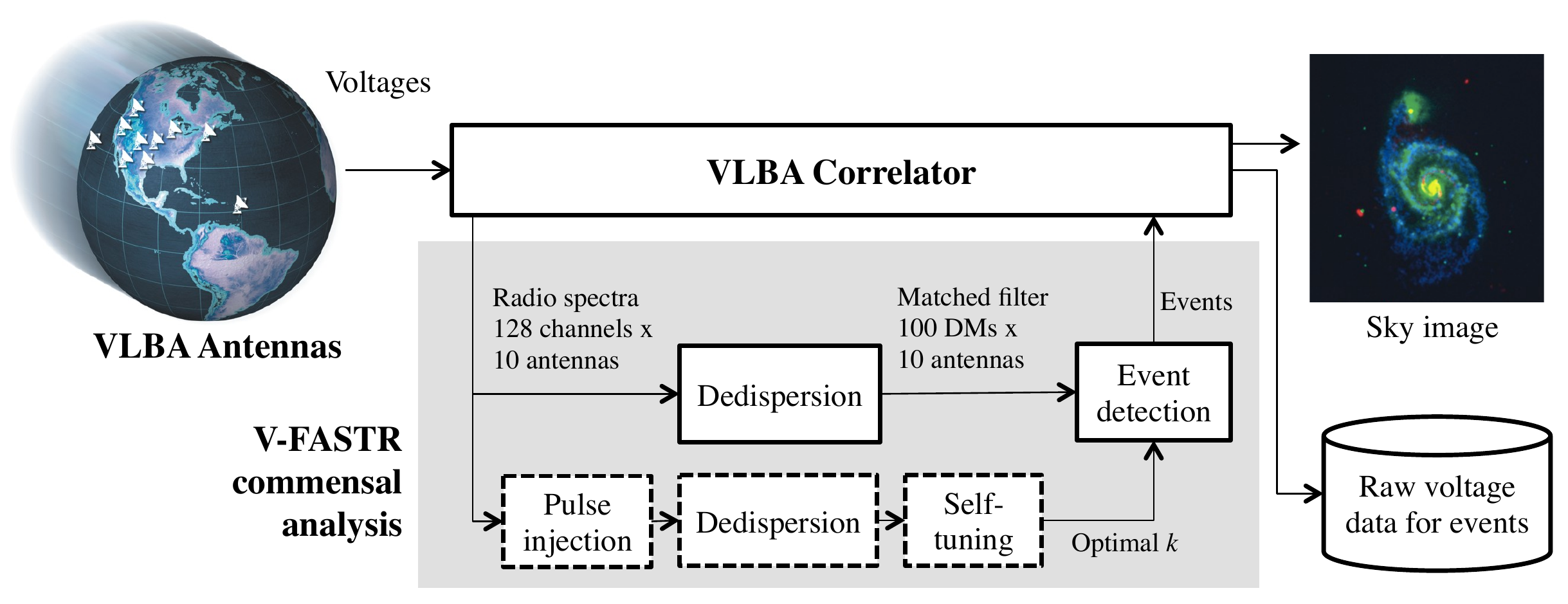}
   \caption{V-FASTR design and interfaces with the VLBA Correlator. The dashed boxes are the adaptive self-tuning system, which runs in alongside the main detection pipeline. Image Credit: NRAO/IAU}
\label{fig:architecture}
\end{figure}

V-FASTR must identify any transients in the short interval before the disks are erased.  The original raw voltage data is important for validation because it retains the phase information of the signal, allowing for vital post-analyses such as imaging (finding the physical position of a pulse on the sky), \edit{coherent summing (combining phased signals to improve sensitivity), and coherent dedispersion (removing dispersive smoothing in the voltage domain).}  Unfortunately, voltage data is so voluminous that it is impossible to archive more than a small fraction.  The V-FASTR commensal detection system \cite{Wayth2011} solves this by searching for events in real time and saving only \edit{ the most promising data}.  \edit{It acts as a plugin to the DiFX software correlator \cite{Deller2007}, which we have modified to broadcast antenna autocorrelations over a secondary network.  In this mode the correlator serves as a spectrometer measuring the channelized, squared, and accumulated power at each antenna across 32 frequency channels.  The integration time interval is approximately $1$ ms. V-FASTR searches this data stream and notifies the correlator of any detected events.  At the end of the correlation process the system saves a copy of these candidate segments to long-term storage for more thorough analysis.}  Figure \ref{fig:architecture} illustrates the V-FASTR system design.   \edit{The commensal analysis runs on a rack-mounted computer with six modern processors but typically uses only a fraction of these resources}.  We describe each software component in greater detail below.

\subsection{Dedispersion search}

The first stage of the V-FASTR analysis is a {\it dedispersion search}, effectively a matched filter search based on knowledge of how signals propagate through the interstellar medium.  It corrects for the distortion caused by free electrons in the space between the source and the receiver.  Following \cite{Lyne1998}, a broadband EM pulse experiences a frequency-dependent delay:
\begin{equation}
t_2 - t_1 = 4.15{\rm ms} ~{\rm DM} [(\nu_1 / {\rm GHz})^{-2}) - (\nu_2 / {\rm GHz})^{-2})] \label{eqn:dispersion}
\end{equation}
\noindent The delay in observed signal arrival between two frequency channels $\nu_1$ and $\nu_2$ is dependent on the bandwidth, the center frequency, and the Dispersion Measure (DM) of the source.  This last value is related to distance and the density of the interstellar medium.  The DM is not known for a new source.  Consequently, the V-FASTR system performs a {\it dedispersion search} over many candidate DMs \cite{Battacharya1998}.  For each timestep it tests a range of possible dispersion measures, equivalent to applying a bank of matched filters with appropriate frequency/time profiles.  Dedispersion searching has proven effective for the detection of pulsars and other astronomical phenomena.  

\begin{figure}[htb]
\includegraphics[width=0.8\textwidth,angle=0]{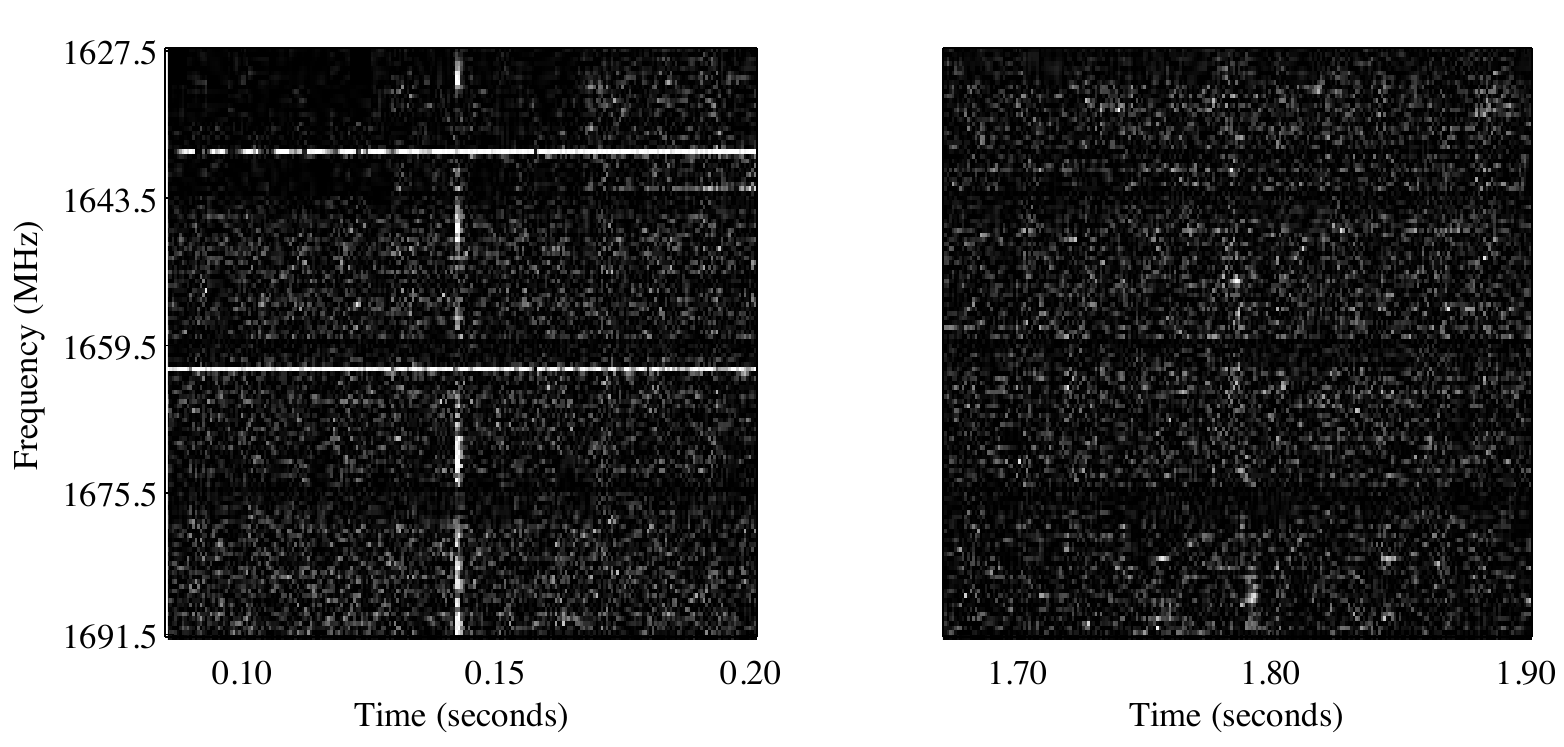}
   \caption{Left: Examples of radio interference, including channelized, momentary, and band-level changes.  Right: A single pulse from pulsar J1919+0021, with a classic dispersed pulse profile.  This plot uses \edit{an incoherent} sum across all VLBA stations to improve contrast; in reality the dedispersion search analyzes each station's data stream separately. The times shown here are consistent with the dedispersed time series shown later in Figure \ref{fig:psr1919}.}
\label{fig:pulse}
\end{figure}

A dedispersion search can also help separate genuine astronomical signals from Radio Frequency Interference (RFI).  Local interference typically manifests as a broadband signal with no dispersion, since the pulse originates locally and all frequencies arrive simultaneously.  Figure \ref{fig:pulse} (Left) is typical of the input data arriving to the V-FASTR system from the correlator.  This dynamic spectrum image displays the data as a matrix of signal powers channelized into discrete time and frequency bins.   The pixel intensity shows observed power, the accumulated squared voltage measured by a station.  This segment contains several kinds of local interference such as narrow-band RFI limited to a few frequency channels and a bright momentary broadband signal near $0.14$ seconds.  The panel at right shows a real astrophysical event: a single pulse from the pulsar J1919+0021, which has an actual DM of 90.315.  Note that the dispersion bends the pulse slightly; the lower frequency components experience greater dispersion delay.  Searching over many DMs transforms the initial spectral data into a vector of approximately 100 matched filter scores - one per DM searched - for each of 10 antennas.   

\subsection{Robust detection of transients}

The VLBA was \edit{designed to have low system temperatures and good gain stability over long timescales.  It is conventionally used in a low time resolution regime.  In contrast, V-FASTR measures signals at high time resolution and experiences additional interference} due to momentary RFI, background signal discontinuities, varying gains, and other artifacts.  We must account for these effects in order to limit the number of false detections.

\begin{figure}[ht]
\centerline{\includegraphics[width=0.7\linewidth]{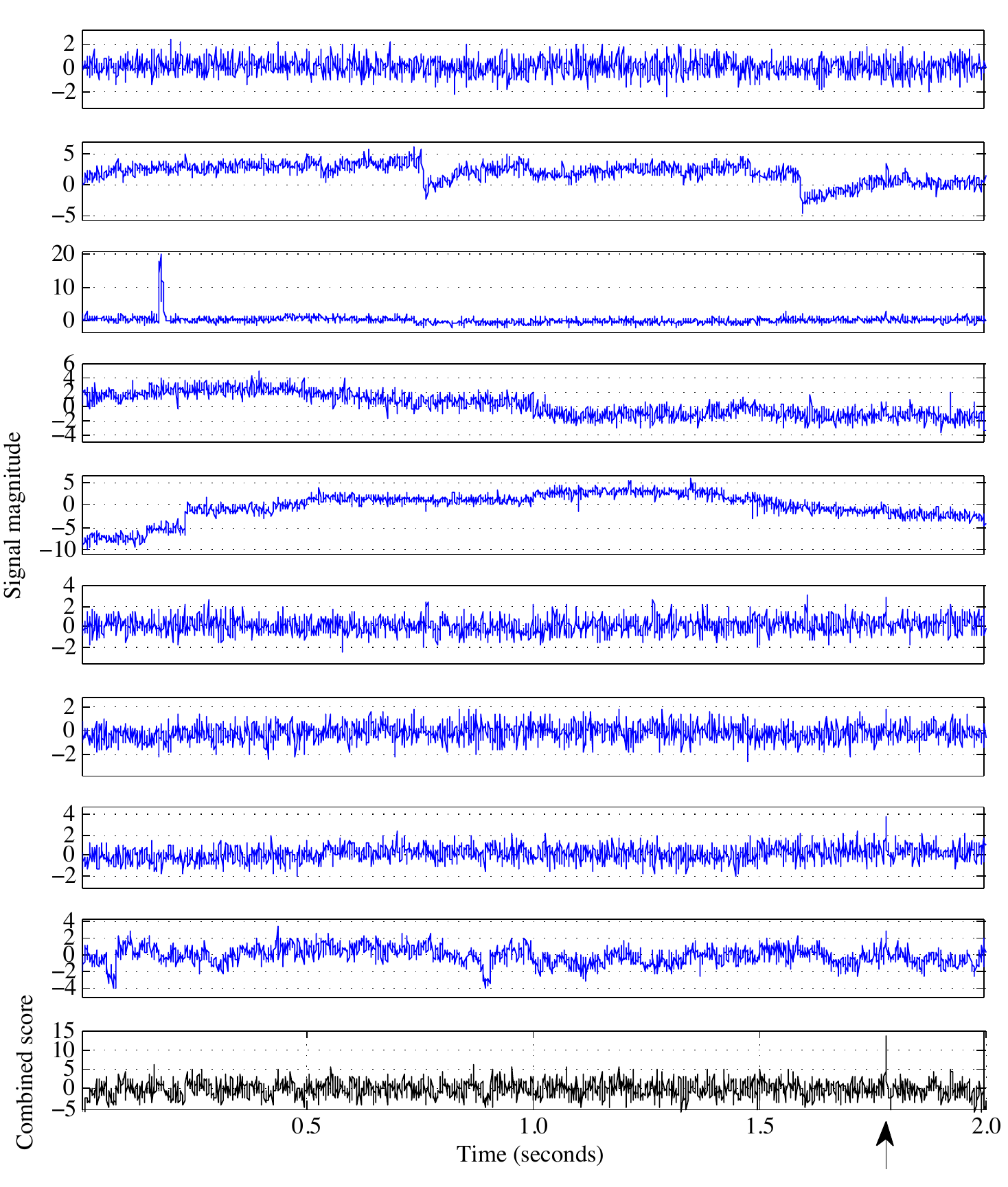}}
\caption{Detection of a single pulse from pulsar J1919+0021.  \edit{The top 9 panels show dedispersed time series from different VLBA antennas.  Non-astrophysical artifacts and interference events appear at several stations. \edit{Such differences are often caused by RFI but could also be caused by maintenance issues, weather, and the source position above each station's horizon}.  The bottom panel shows the scalar output of the adaptive detection score. An arrow indicates the event - the only apparent pulse from this source. It was missed by the standard detection approach but successfully found by the V-FASTR system.}}
\label{fig:psr1919}
\end{figure}

We exploit the fact that the stations are geographically separated, so interference only appears at a single station while true astrophysical signals are correlated across many stations \cite{Thompson2011}.  \edit{The top nine rows of Figure \ref{fig:psr1919} show} the signal from each antenna after dedispersion to a DM of 90. An arrow shows a real astronomical pulse from the pulsar J1919+0021.  The pulse appears in multiple stations simultaneously, indicating that it is \edit{isolated on the celestial sphere}.  Because we do not know the correct DM in advance, many tens of these time series are produced simultaneously and the system must analyze them all.  This provides a real-valued input vector having dimensionality of order one hundred.  The task amounts to a multivariate pattern recognition problem with millisecond time requirements, a nonstationary background distribution, and a very low tolerance for false positives ($10^{-5}$ false positives per timestep).  

We represent each sample in the dedispersed time series as a multi-element vector $\x \in \mathbb{R}^{d\times n}$ where $d$ is the number of dispersion measures and $n$ the number of stations.   We can formalize the detection rule as a function $f_k(\x)$ that maps the dedispersed time series to a scalar score which is high for true transients but low for interference.  \edit{We anticipate that any astrophysical fast transients will be very rare.  For instance, \cite{Madau1998,Siemion2011} estimate the rate in a local 1 Gpc$^3$ volume to be one per square degree of sky per 1000 hours.}  For the VLBA \edit{primary beam} field of view of 0.27 deg$^2$ and $1$ ms temporal quantization this amounts to a target/background fraction \edit{of approxmately} $1\times 10^{-10}$ \cite{Wayth2012,Wayth2011}.  

The conventional approach to signal detection in VLBA data is to sum the signals from all antennas.  As described in previous work \cite{Thompson2011}, the V-FASTR system favors a simple \emph{robust estimator} that excises one or more extreme values and averages the rest.  \edit{First the data stream from each station is noise-whitened, standardized to have zero mean and unit standard deviation, dedispersed, and finally filtered using a high-pass filter with a $100$ ms diameter. We then apply a separate matched filter for each DM and store the result as a multivariate time series.}  Suppose $\x_{id}$ is the matched filter result from station $i$ at dispersion measure $d$.  We sort the antennas in order of increasing signal strength for each DM and sum all but the most extreme $k$ station signals to make a combined signal.  The maximum response over all DMs becomes the overall score for that timestep:   
\begin{equation}
{\rm Detect~ if} ~ f_k(\x) \geq \tau \quad {\rm for} \quad f_k(\x) = \displaystyle\sup_{d} \frac{1}{(n-k)}\displaystyle\sum_{i=1}^{n-k}\x_{id}\\
\end{equation}
\noindent This is most appropriate if the stations provide independent and identically distributed estimates of some common source perturbed by occasional sporadic noise.  Operationally we define $\tau$ in terms of a Signal to Noise Ratio (SNR) given in standard deviations from the mean. 

The best value of $k$ depends on noise statistics.  V-FASTR uses incoherent addition that ignores phase information, so the measured SNR for a real event approximately follows the relationship $\sqrt{n-k}$ and excising too many antennas reduces sensitivity.  However, excising too few extreme values increases the false positive rate.  The optimal tradeoff depends on background noise and the number of antennas participating in the observation.  

We formalize the expected loss for a particular value of $k$ using Bayesian decision theory.  Each timestep is associated with a background event $y_0$ or a transient event $y_1$, with transient events much rarer than the background. We posit a misclassification cost $c_0$ for a false positive detection, and $c_1$ for missing a real transient event.  We optimize $k$ and $\tau$ to minimize Bayes' loss, i.e. the expected loss considering \edit{both error probabilities}.
\begin{eqnarray}
L(\tau,k) &=& c_1~p(f_k(\x)<\tau~|y_1) ~p(y_1) + c_0~p(f_k(\x)\geq\tau~|y_0)~p(y_0) \\
& =& \alpha~p(f_k(\x)<\tau~|y_1) + p(f_k(\x)\geq\tau~|y_0) \label{eqn:loss}
\end{eqnarray}
\noindent The $\alpha$ term is a constant that incorporates the prior probability of each event as well as the relative costs of misclassification. Even positing a very large weight for $c_1$, the class imbalance is much larger and we can assume $\alpha \ll 1$.  

\subsection{Adaptive self-tuning for RFI excision}

It is challenging to construct a universal detection rule because Equation \ref{eqn:loss} depends on noise properties that are unknown in advance and difficult to model.  Instead we use the {\it data stream itself} for an empirical nonparametric estimate of the loss function $L(\tau,k)$.  Specifically, we use Equation \ref{eqn:dispersion} to construct synthetic pulses at known locations and dispersions, add these signals to the input stream and then attempt to recover them using a range of parameterizations.  We set $\tau$ to be the minimum value that does not produce any significant false positives (e.g., very low-$\alpha$ regimes).  We use the number of synthetic pulses recovered prior to the first $m$ false positives, for low $m$, as a figure of merit to select $k$.  
 
During operations the commensal system buffers 10,000 timesteps of spectrometer data from the correlator and retrains itself on the preceding batch prior to  detection.  It injects 200 dispersed pulses at SNRs from $5$ to $9$ and DMs from $10$ to $50$.  \edit{This range} does not span the full dedispersion search, but it allows the system to fit more pulses into the same time interval.  \edit{This parallel stream is then dedispersed and analyzed with all possible excision levels}.  \edit{We compute the detection figure of merit permitting a small number $m$ of real positives (typically 20) so that single, isolated real pulses do not significantly affect the result.  Typically the system excises a single antenna but can remove up to four in scans having very high interference}.  \edit{ In Figure \ref{fig:psr1919} the system elected to excise two antennas, resulting in the combined detection score shown in the bottom panel.}

\section{Performance}

This section evaluates the adaptive system's performance. The VLBA occasionally observes known pulsars during its normal operations, which provides a blind sensitivity test.   To date, the commensal system has detected hundreds of events from seven different pulsars (Table \ref{tab:pulsardetections}).  In contrast, the conventional non-adaptive approach of summing signals has failed to detect the two weakest pulsars (J1919+0021 and J0146+5922).  Figure \ref{fig:psr1919} shows the J1919+0021 detection.  

\begin{table}[ht]
\begin{centering}
\begin{tabular}{cccc} 
Pulsar name & Max SNR & DM & On-pulse flux density (Jy)\\ 
\hline 
J0332+5434 & $>$50 & 26 &14.0\\ 
J1136+1551 & 20 &5 & 1.0\\ 
J0826+2637 & 19 &20 &0.5\\ 
J1935+1616 & 12 &158& 1.0\\ 
\edit{J01645-0317} &  \edit{10} & \edit{36} &  \edit{0.5} \\
J0157+6212 & 10 &30& 0.14\\ 
J1919+0021 & 9 & 90 &0.05 \\
J0147+5922 & 9 & 40 & 0.13\\
\edit{J0953+0755} & \edit{8} & \edit{3} & \edit{2.0}\\
\hline
\end{tabular}
\end{centering}
\caption{Known pulsars detected serendipitously by V-FASTR.}
\label{tab:pulsardetections}
\end{table}
 
\begin{figure}[ht]
\centerline{\includegraphics[width=0.6\linewidth]{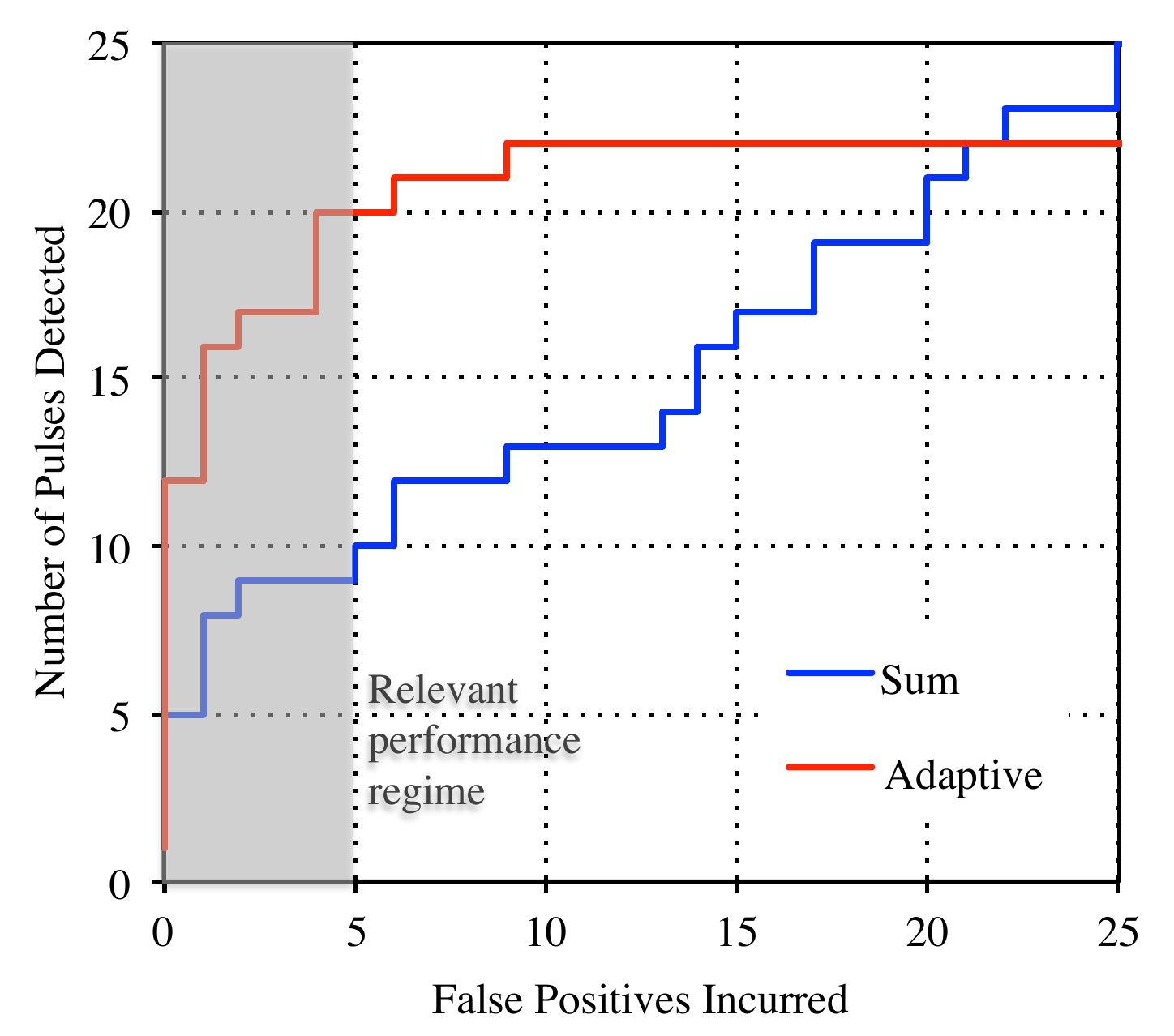}}
\caption{Detection performance for pulsar J0826+2637 by adaptive and non-adaptive (summing) approaches.}
\label{fig:roc}
\end{figure}

We also compared detection results for observations containing many pulses.  Figure \edit{\ref{fig:roc}} shows a Receiver Operating Characteristic (ROC) curve for a scan containing over 20 pulses from J0826+2637.   Not all of the curve is relevant to our task \edit{since} the correlator can only archive a small \edit{number} of events in the time interval between detection and disk erasure.  Typically it saves at most four or five events with a priority ordering determined by their signal strength.  Consequently the settings with fewer than five false positives, shaded in gray, are most important.  The adaptive approach finds a far greater number of real astrophysical events meaning it successfully prioritizes weak real pulses over RFI.  We have tuned the adaptive system for maximum precision in the critical low-false-positive regime. 

\begin{figure}[ht]
\centerline{\includegraphics[width=0.5\linewidth]{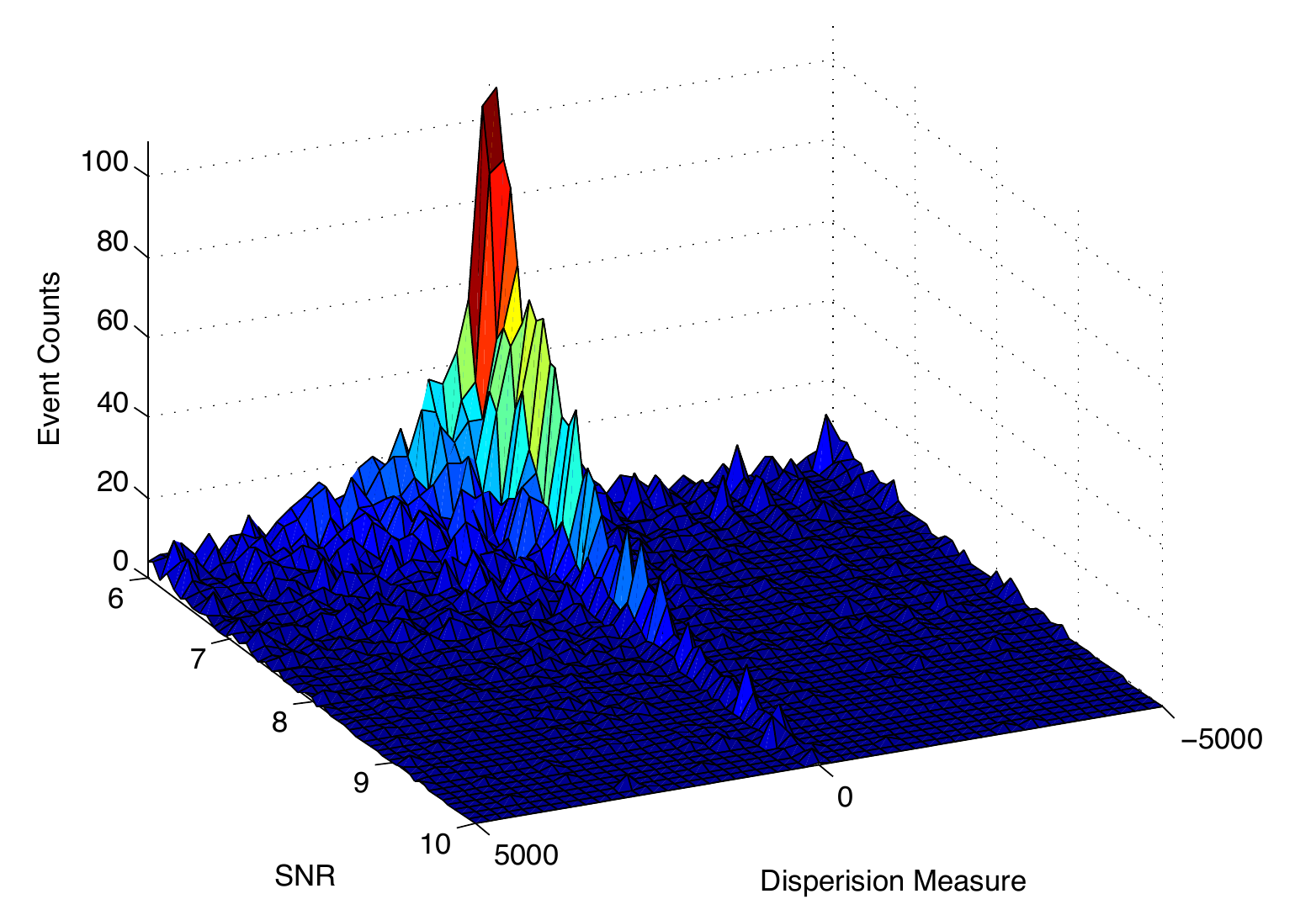}\includegraphics[width=0.5\linewidth]{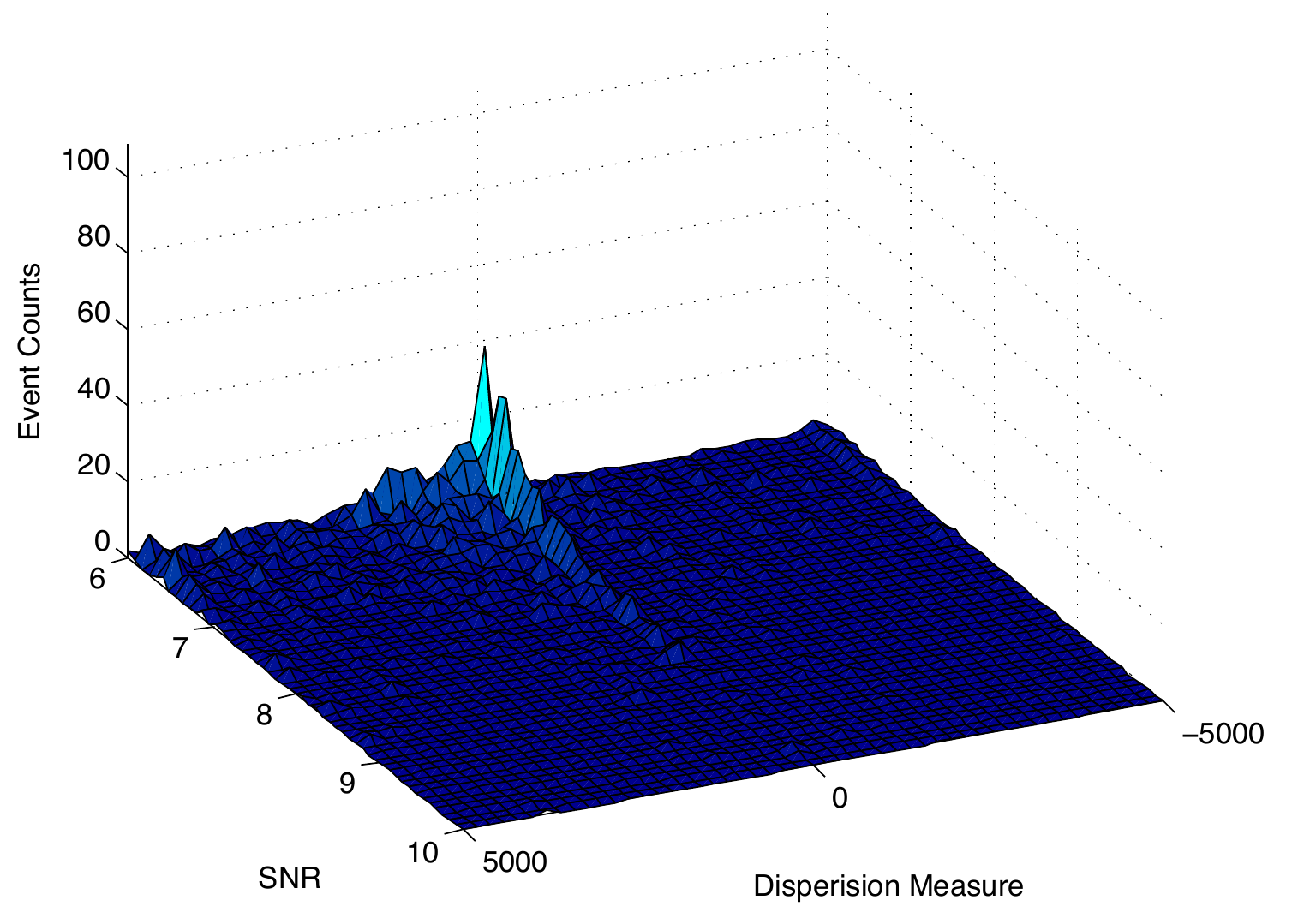}}
\caption{The total count of false positive detections for the month of April 2012.   The vertical axis shows the number of detections while the horizontal axes show SNR and DM values.  Left: False positives detected by the standard summing approach. Right: False positive detected by the V-FASTR adaptive method.}
\label{fig:totalcounts}
\end{figure}

A major bottleneck to sustained commensal analyses is the limited analyst time available to review any candidate detections.  The V-FASTR policy is to manually review all detections with a signal-to-noise ratio above 7. {\it Post-hoc} analyses of V-FASTR performance show that the adaptive system has greatly reduced the overall number of these candidates that must be reviewed.  April 2012 provided a good test case since there were no pulsars or other \edit{known} astrophysical transient events being observed.   We applied a standard RFI filter to remove all signals with a DM less than 10.  Any remaining detections in April are false positives.  Figure \ref{fig:totalcounts} shows the number of detections at different SNR thresholds, for both adaptive and non-adaptive approaches. The non-adaptive approach yields an unfeasibly large number of detections for the same SNR threshold.   

Finally, we recorded the excision levels actually selected by the self tuning system during this period.  Several very clean observations at high frequencies resulted in the system choosing $k=0$, tantamount to no RFI protection.  These scans did not produce any false positive events.  For detected events, the  proportions of excision levels in use were 54.7\% (k=1), 23.9\% (k=2), 10.6\% (k=3), and 10.7\% (k=4).  Over $78\%$ of the detections \edit{used} an excision level of just one or two stations.   The larger excision levels appear only during extremely noisy observations or those containing pulsed signals from satellites.  Satellite signals can appear in multiple stations simultaneously, and are currently the largest source of false positives.

\section{Discussion, Conclusions, and Lessons Learned}

V-FASTR performs adaptive detection of fast transients by commensal analysis at the Very Long Baseline Array.   Its ability to retrain automatically through injection of synthetic events has improved reliability over traditional detection approaches.   This system has operated for over one year and has recently placed new limits on the rates of astrophysical phenomena observed at 1.4 GHz and above \cite{Wayth2012}.  This makes it a working example of adaptive pattern recognition in data streams to enable new science results.  V-FASTR continues to accumulate telescope time at multiple frequencies; manageable false positive rates mean that the system is sustainable for the long term.    

The effective composition and coordination of the team has been crucial factor in this successful deployment.  The V-FASTR group combines experts in machine learning with experts in radio astronomy and the VLBA's core processing system.  This fusion of expertise kept the machine learning approaches grounded in the application and operational perspectives.  As a result, the VLBA has adopted V-FASTR for full-time use, establishing the policy that all observing campaigns have their data analyzed by the commensal system by default.  Naturally, any interesting detections made by V-FASTR would immediately result in the involvement of that campaign's Principal Investigator.  The provision of immediate data access to the automated system is a notable policy innovation that paves the way for more commensal automated science concepts.  Other instruments and telescopes have expressed interest in similar systems to aid in their own processing pipelines. 

\section{Acknowledgements}
We thank Peter Hall and J-P Macquart (Curtin University/ICRAR), as well as Dayton Jones, Robert Preston, and Joseph Lazio (Jet Propulsion Laboratory).  The International Centre for Radio Astronomy Research is a Joint Venture between Curtin University and The University of Western Australia, funded by the State Government of Western Australia and the Joint Venture partners. S.J.T is a Western Australian Premiers Research Fellow. R.B.W is supported via the Western Australian Centre of Excellence in Radio Astronomy Science and Engineering.  The National Radio Astronomy Observatory is a facility of the National Science Foundation operated under cooperative agreement by Associated Universities, Inc.  Part of this research was performed at the Jet Propulsion Laboratory, California Institute of Technology, under the Research and Technology Development Strategic Initiative Program, under a contract with the National Aeronautics and Space Administration.  The research was also supported in part by the Keck Institute for Space Studies.  Copyright 2013 California Institute of Technology.  All Rights Reserved.  U.S. Government Support Acknowledged.
\bibliographystyle{latex8}
\bibliography{2012_V-FASTR}
\section{Biographies}

Dr.~David R.~Thompson is a research technologist at the Jet Propulsion Laboratory, California Institute of Technology and a member of IEEE. His research applies machine learning to remote exploration.

Dr.~Sarah Burke Spolaor is a post-doctoral astrophysics researcher at the Jet Propulsion Laboratory, California Institute of Technology. Her research focuses on pulsar astronomy and detecting short-timescale radio transients.

Dr.~Adam T.~Deller is a VENI postdoctoral fellow at ASTRON, the Netherlands Institute for Radio Astronomy.  His research interests include high precision astrometry of compact radio sources and instrumentation development for Very Long Baseline Interferometry.

Dr.~Walid A.~Majid is a senior researcher at the Jet Propulsion Laboratory, California Institute of Technology.  His research involves pulsar astronomy and development of instrumentation and signal processing techniques for radio astronomy.

Divya Palaniswamy is a PhD candidate at Curtin University whose thesis work focuses on high time resolution radio astronomy.

Dr. Steven J.~Tingay is a Professor of Radio Astronomy at Curtin University in Perth, Western Australia, and a Western Australian Premier's Research Fellow.  His research focuses on radio astronomy instrumentation and astrophysics.

Dr.~Kiri L.~Wagstaff is a senior researcher at the Jet Propulsion Laboratory, California Institute of Technology, where she focuses on the use of machine learning methods to investigate science questions using planetary and astronomical data.

Dr.~Randall B.~Wayth is a research scientist at Curtin University in Western Australia, and a member of IEEE. His research interests include radio astronomy instrumentation and astrophysics.

\end{document}